\documentclass[runningheads]{llncs}
\usepackage{graphicx}

\begin{document}
\title{Speaker Adaptation with Continuous Vocoder-based DNN-TTS}
%
%
\author{Ali Raheem Mandeel\inst{1}\and
 Mohammed Salah Al-Radhi \inst{1}\and
Tamás Gábor Csapó\inst{1,2}}
\authorrunning{Ali Raheem Mandeel et al.}
%
\institute{Department of Telecommunication and Media Informatics,\\ Budapest University of Technology and Economics, Budapest, Hungary \and
MTA-ELTE Lendület Lingual Articulation Research Group, Budapest, Hungary
\email{aliraheem.mandeel@edu.bme.hu, \{malradhi,csapot\}@tmit.bme.hu}\\}

\maketitle              
\begin{abstract}
Traditional vocoder-based statistical parametric speech synthesis can be advantageous in applications that require low computational complexity. Recent neural vocoders, which can produce high naturalness, still cannot fulfill the requirement of being real-time during synthesis. In this paper, we experiment with our earlier continuous vocoder, in which the excitation is modeled with two one-dimensional parameters: continuous F0 and Maximum Voiced Frequency. We show on the data of 9 speakers that an average voice can be trained for DNN-TTS, and speaker adaptation is feasible 400 utterances (about 14 minutes). Objective experiments support that the quality of speaker adaptation with Continuous Vocoder-based DNN-TTS is similar to the quality of the speaker adaptation with a WORLD Vocoder-based baseline.

\end{abstract}
\section{Introduction}
\begin{figure}
  \centering
  \includegraphics[width=\textwidth]{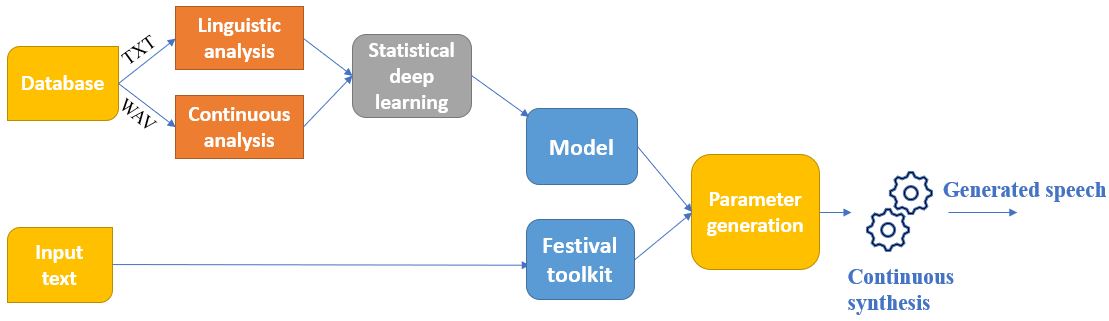}
  \caption{Schematic diagram of statistical parametric speech synthesis.}
  \label{fig:speech_sysnthesis}
\end{figure}

Speech processing has brought the attention of researchers and industry as well during the last decades. The rapid advancement in digital technology has led to a wider  variety of speech processing functions (such as speech recognition, speech synthesis, dialogue control, and so on) is becoming a central mechanism for creating a human-computer communication interaction. Speech could be used in a variety of industries, including machine control using natural speech rather than written commands, surveillance, healthcare, and the Internet of Things. The technique of converting text into artificial speech is known as speech synthesis which is opposite to Automatic speech recognition (ASR) ~\cite{habeeb2020ensemble}. Nowadays, all efforts are devoted to producing a sound similar to the natural sound.\\
Text-to-speech (TTS) synthesis in the state-of-the-art is based on computational parametric methods. Its benefit is the availability of speaker adaptation techniques, which allow for the creation of a unique voice based on any target speaker. The speech signal is broken down into parameters expressing excitation fundamental frequency (F0) and speech spectrum in the parametric model, which are then loaded into a machine learning system. In the synthesis, the parameter chains are reconverted to voice signal using rebuild approaches (excitation models, vocoders) after the mathematical model has been learned on the training data. When it comes to speech synthesis solutions, if the system is configured for real-time low latency use, it is possible to produce audible speech output with a live person. An example is the use at government websites (e.g. the Hungarian Chatbot using TTS techniques at https://ugyfelkapu.gov.hu/), which should be prepared for extremely high loads, and therefore, real-time synthesis of speech is extremely important.  Figure \ref{fig:speech_sysnthesis} shows the statistical parametric speech synthesis system’s basic components. \\
While TTS systems are now intelligible, existing parametric techniques do not allow for absolute naturalness in real-time systems, and there is still potential for advancement in terms of being as similar to human speech as possible. While there are vocoding approaches that produce natural-sounding synthesized speech (e.g. STRAIGHT and WORLD), they are usually computationally costly and hence unsuitable for real-time execution. With minimal training and adaptation data, a speech synthesis system should have the ability to produce the voice of any speaker. A substantial advantage of SPSS (statistical parametric speech synthesis) above unit-selection speech synthesis is its versatility in altering speaker traits, emotions, and speaking styles, which is due to the potential performance of speaker adaptation ~\cite{zen2009statistical}. Additionally, the efficiency of DNN-based adaptation is superior to HMM-based adaptation ~\cite{wu2015study}. Meanwhile, Deep neural networks (DNN) have significantly improved SPSS according to ~\cite{ling2013modeling}, ~\cite{wu2015deep}, and ~\cite{hashimoto2015effect}.\\ 
The capability to generate new voices utilizing just a limited quantity of adaptation data from an objective speaker can be achieved using adaptation techniques ~\cite{lanchantin2014multiple}. Because of its function in linking linguistic and acoustic features, the acoustic model's constraint reduces the level of speech naturalness. The continuous vocoder outperforms other traditional vocoders with discontinuous F0 because the vocoder parameters are simple ~\cite{al2018continuous}. Our novel contribution in this work to use the continuous vocoder with speaker adaptation in Merlin (a speech synthesis toolkit that uses neural networks to create speech)~\cite{wu2016merlin}. We show on the data of 9 speakers that an average voice can be trained, and speaker adaptation is feasible with limited data of 400 utterances (about 14 minutes). Objective experiments support that the quality is similar to the WORLD-based baseline.\\  
This paper is organized in the following sections: An overview of the related scientific papers, including the novel methods in speech adaptation based text-to-speech synthesis. In the second section, the methodology was described and the design, tools, and dataset were clarified. After that, in section three, the results were explained. Finally, the conclusion was mentioned in section four. \\ 

\section{Related work}
 Until 2013, voice synthesis was dominated by unit-selection synthesis and Hidden Markov model (HMM)-based voice synthesis. There are many fusion approaches between unit selection and statistical parametric synthesis. A review presents the main technique of statistical parametric speech synthesis ~\cite{black2007statistical}. According to the authors, the combination of the two approaches will carry the advantages of both and eliminate some drawbacks. The adaptability and controllability of statistical parametric speech synthesis (SPSS) over unit-selection speech synthesis is a significant benefit. Speech synthesis employing a hidden Markov model (HMM) has many benefits, including the ability to alter the speaker's identity, emotion, and speech style ~\cite{tokuda2013speech}. This technique, however, is a long way from natural speech.~\cite{zen2014deep} used a mixture density output layer to solve the inability of DNNs to estimate variances in SPSS (MDNs). It will calculate maximum probability density functions above actual-valued production features based on the equivalent input characteristics. The findings indicate that better predicting acoustic features improved the naturalism of synthesized voice.\\
~\cite{beskow2016hybrid} used a fusion method to modeling the speech signal to obtain more normal synthesized speech than the purely statistical version. Via spectrogram kernel filtering, the speech was decomposed into a harmonic and a noise burst component. A vocoder and statistical parameter generation are used to model the harmonic component, while concatenation is used to model the burst part. The final synthesized waveform is created by combining the dual channels.\\
A new method for using sinusoidal vocoders in DNN-based statistical parametric speech synthesis (SPSS) was introduced by ~\cite{hu2015fusion}. At the statistical modeling and synthesis stages, the system uses sinusoids as a simple parameterization (DIR) or an intermediate spectral parameterization (INT). Both approaches lead to improving modeling accuracy, according to the findings. Other objective functions can be chosen and tailored for the fusion process to increase the perceptual outcome in the potential work of this research.\\
A comprehensive and experimental study of speaker adaptation with deep neural network (DNN) based speech synthesis was conducted by ~\cite{wu2015study}. Adapting to a neural network can be done in three ways. The first method involves performing feature space transformations, the second entails augmenting speaker-particular features as feedback to neural networks, and the third method includes model adaptation. The future work of this study to consternate on more precise modeling of the source-filter interaction using the DNN-based approach, as this model has been shown to be insufficient for significantly improving segmental speech efficiency.\\
In statistical parametric speech synthesis,~\cite{airaksinen2018comparison} compared STRAIGHT, glottal, and sinusoidal vocoding. Vocoder accuracy was evaluated in a statistical parametric speech synthesis system utilizing four distinct voices in the form of assessment-synthesis in addition to TTS. The findings suggest that the superiority of vocoder-produced speech is influenced by the voice used. Furthermore, using the glottal vocoder GlottDNN and a male voice with a limited range of expression, the single best-rated TTS system was obtained. Moreover, the sinusoidal vocoder PML (pulse model in log-domain) has the highest aggregated result in all of the tests conducted.~\cite{agiomyrgiannakis2015vocaine} presented a new vocoder synthesizer (Vocaine) with an Amplitude Modulated-Frequency Modulated speech model. Vocaine  Outperform STRAIGHT unit-selection speech synthesis in computation and match them in performance. Maximum Voiced Frequency is employed for combing voiced and unvoiced excitation by a residual-based vocoder to remove artifacts created by creaky speech in the SPSS ~\cite{csapo2015residual}. The vocoder's parameters are all continuous, so Multi-Space Distribution isn't needed when training the HMMs.\\
~\cite{hu2013experimental} used the same speech data and controlled laboratory conditions to compare various vocoders in a large-scale listening exercise. The responses were then studied and visualized using K-means clustering and multidimensional scaling, as well as the interaction between the various vocoders. The Vocoders are used to convert acoustic parameters into human voices and vice versa. In text-to-speech synthesis, the vocoders are the most critical part. A Vocoder acts as both an analyzer and a synthesizer. It analyzes speech by transforming the waveform into a series of parameters that reflect the vocal-folds excitation signal and filtering the excitation signal with a vocal-tract filter transfer tool. Otherwise, it reconstructs the initial speech signal from the used parameters. There are several types of vocoders such as STRAIGHT, WORLD, Mel - generalised cepstral vocoder, adaptive harmonic model, Glottal vocoder, and Harmonic model, etc. ~\cite{hu2013experimental}. The continuous vocoder, despite being an alternative solution, hasn't been tested in speaker adaptation yet.\\

\section{Methodology}
\subsection{Speech corpora}
The studies were carried out using the VCTK-Corpus ~\cite{bakhturina2021hi}. This CSTR VCTK Corpus contains speech data from 109 (with 47 of them being male) English speakers who speak with a variety of accents. Approximately 400 sentences are read by each speaker. Over 9 speakers, we ran the average speech model (AVM) (six females and three males). The waveforms in this database are re-sampled to 16 kHz.

\subsection{Continuous Vocoder}
In the Continuous vocoder ~\cite{csapo2016modeling}, the fundamental frequency (F0) is determined on the input waveforms during the analysis stage. Following this, the MVF parameter is determined from the speech signal. The speech signal is then subjected to gamma=-1/3 with 24-order Mel-Generalized Cepstral analysis (MGC) and alpha = 0.42 as the next move. The frameshift is 5ms in both measures. Furthermore, in the voiced portions of the opposite filtered residual signal, the Glottal Closure Instant (GCI) algorithm is utilized to discover the glottal period limitations of particular periods. A PCA residual is constructed from these F0 cycles, which will be used in the synthesis process. White noise is used during the synthesis process because the frequencies are greater than the MVF value. To eliminate any residual buzziness and reduce the noise portion, applying a time-domain envelope to the unvoiced segments is proposed as a method of modeling unvoiced sounds ~\cite{al2020continuous} and  ~\cite{al2017time}. The vocoders use different parameters as can see in the table \ref{tab:vocoders}.\\

\begin{table}
\caption{The vocoders types and their parameters }
  \label{tab:vocoders}
  \centering
\begin{tabular}{|l|l|}
\hline
Vocoder name                       & Parameters per frame                                                                                                    \\ \hline
Continuous                         & \begin{tabular}[c]{@{}l@{}}F0: F0: 1 + MVF: 1 + MGC: 24\end{tabular}                                                    \\ \hline
STRAIGHT                           & \begin{tabular}[c]{@{}l@{}}F0: 1 + Aperiodicity: 1024\\  + Spectrum: 1024\end{tabular}                                  \\ \hline
WORLD                              & \begin{tabular}[c]{@{}l@{}}F0: 1 + Band aperiodicity:\\ 5 + MGC: 60\end{tabular}                                        \\ \hline
Mel - generalised cepstral vocoder & \begin{tabular}[c]{@{}l@{}}MGC: 24 + F0: 1 , Pulse\\ plus noise excitation\end{tabular}                                 \\ \hline
Glottal vocoder                    & \begin{tabular}[c]{@{}l@{}}F0:1, Energy:1, HNR: 5,\\ Source LSF: 10, Vocal tract \\ LSF: 30, natural pulse\end{tabular} \\ \hline
Harmonic model                     & \begin{tabular}[c]{@{}l@{}}2*k harmonics + F0:1, \\ Harmonic excitation\end{tabular}                                    \\ \hline
Adaptive harmonic model            & \begin{tabular}[c]{@{}l@{}}2*k + F0:1, Harmonic\\  excitation\end{tabular}                                              \\ \hline
\end{tabular}
\end{table}

\subsection{Build an average voice model (AVM)} 

Over the range of 9 speakers, we created the average voice model (AVM) (six females and three males). We set up the data and directories, prepared the labels, prepared acoustic features (MGC, MVF, and lf0) using the continuous vocoder. Then we trained duration and acoustic models. Finally, we synthesized speech. 
We used a Feedforward neural network of 6 hidden layers with tangent hyperbilic units (1024 neurons, 265 batch size, sgd optimizer, and 25 epochs). This neural network is the most basic kind of networks. This architecture is known as a Deep Neural Network (DNN) when there are enough layers. Several layers of hidden units, each performing a nonlinear operation, are used to estimate the output from the signal. It uses linguistic features as input data to estimate vocoder parameters.\\ 

\subsection{Adapt the AVM for the adapt speaker}  

We did the adaptation with multiple speakers (2 females, and 2 males). We adapted the four speakers (p234, p236, p237, and p247) using a Feedforward neural network with the continuous vocoder. MGC, MVf, and F0 are the parameters used in this vocoder. The adaptation data for each speaker is 400 utterances (about 14 minutes). 

\section{Results}
Objective and subjective analyses were performed in order to meet our purposes and to validate the feasibility of the proposed process. We independently checked our continuous vocoder parameters with speaker adaptation in deep neural networks using MCD, and F0-CORR, and spectrogram analysis in the objective evaluation. Moreover, a subjective listening experiment was used to test them.

\subsection{Objective evaluation}

\begin{enumerate}
\item MCD(dB):
Distortion of Mel Cepstral measurement 60-dimensional coefficients for each training model were measured. The efficiency of the continuous vocoder systems against the WORLD baseline scheme could be concluded based on these findings.

\begin{equation}
    M C D=\frac{1}{N} \sum_{j=1}^{N} \sqrt{\sum_{i=1}^{K}\left(x_{i, j}-y_{i, j}\right)^{2}}
\end{equation}
x and y are the actual and synthesized voice signals of ith cepstral coefficients. The MCD errors on the validation (dev) and test sets results are described in the table \ref{tab:t2}. It seems that the WORLD vocoder has a bit better (i.e, lower MCD) than continuous vocoder. 
 
\begin{table}[]
 \caption{MCD errors on the dev/test sets.}
  \label{tab:t2}
   \centering
\begin{tabular}{|c|c|c|}
\hline
Spkr & WORLD vocoder & Continuous vocoder \\ \hline
P234 & 5.309 / 5.251 & 5.440 / 5.434      \\ \hline
P236 & 5.540 / 5.400 & 5.795 / 5.547      \\ \hline
P237 & 5.152 / 5.072 & 5.370 / 5.240      \\ \hline
P247 & 5.347 / 5.335 & 5.502 / 5.483      \\ \hline
\end{tabular}
\end{table}
\item F0-CORR: 
the correlation is a reflection of how closely reference and produced data are related (linearly related). Overall, Frame by frame, a measurement is made. F0-CORR shows good results for continuous vocoder as the values reaching 1 and higher than the baseline vocoder (table \ref{tab:t3}).
\begin{equation}
    F0-CORR=\frac{\sum_{i=1}^{n}\left(x_{i}-\overline{\mathrm{x}}\right)\left(y_{i}-\overline{\mathrm{y}}\right)}{\sqrt{\sum_{i=1}^{n}\left(x_{i}-\overline{\mathrm{x}}\right)^{2}} \sqrt{\sum_{i=1}^{n}\left(y_{i}-\overline{\mathrm{y}}\right)^{2}}}
\end{equation}
\begin{table}[]
\caption{F0-CORR on the dev/test sets.}
  \label{tab:t3}
   \centering
\begin{tabular}{|c|c|c|}
\hline
Spkr & WORLD vocoder & Continuous   vocoder \\ \hline
P234 & 0.456 / 0.511 & 0.730 / 0.755        \\ \hline
P236 & 0.607 / 0.596 & 0.480 / 0.588        \\ \hline
P237 & 0.582 / 0.553 & 0.760 / 0.721        \\ \hline
P247 & 0.608 / 0.655 & 0.766 / 0.682        \\ \hline
\end{tabular}
\end{table}

\item  Spectrogram Analysis:
the visual representation of synthesized sounds by both vocoders has been shown in figure \ref{fig:spectrogram}. We show four spectrogram plots of two speakers (a female and a male) with two vocoders (continuous and WORLD) who read the sentence "I had faith in them.". As can be seen in the spectrograms, the continuous vocoder separates the voiced and unvoiced frequencies of speech, according to the MVF parameter (i.e., the spectral content below MVF is voiced, and above MVF is unvoiced). The WORLD vocoder handles the voicing via band aperiodicities, therefore, such a strong separating curve is not visible in the spectrogram.

\begin{figure}
  \centering
  \includegraphics[width=\textwidth]{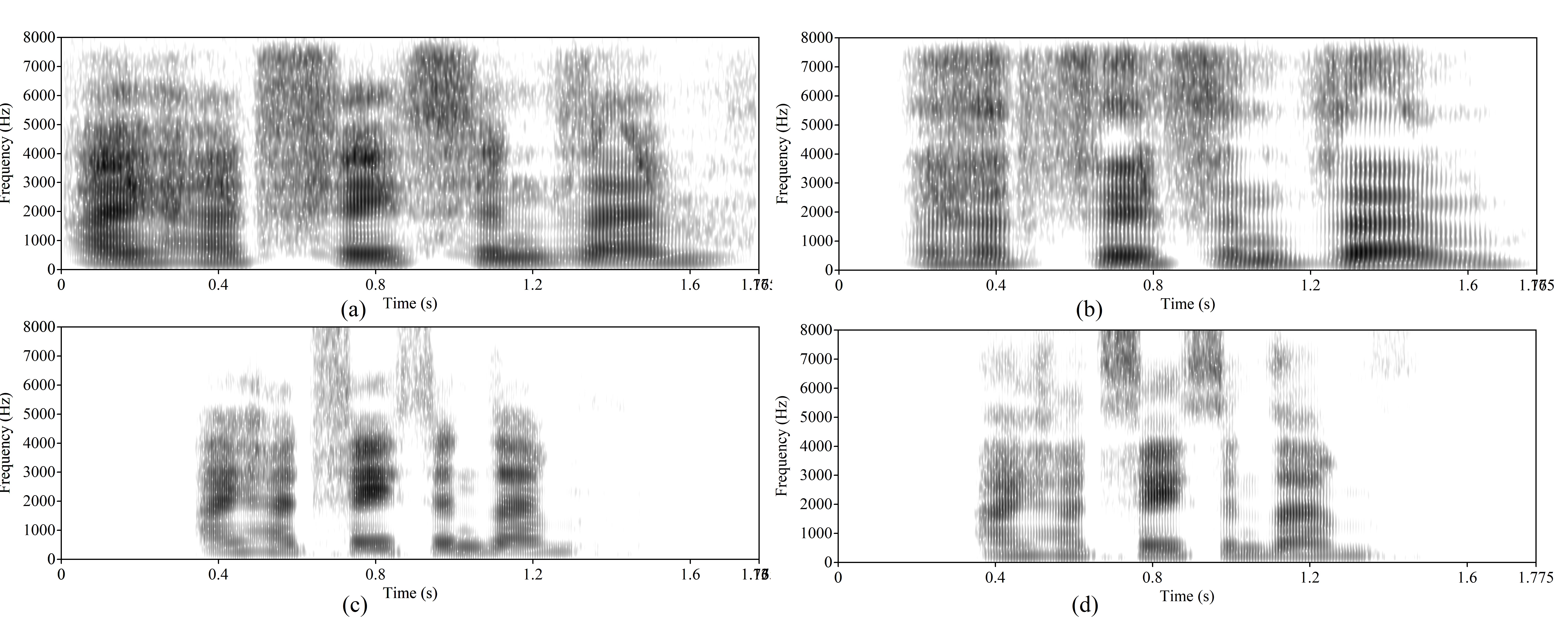}
  \caption{The spectrogram plot of four synthesised sounds (a) a female voice using continuous vocoder (b) a male voice using continuous vocoder (c) a female voice using WORLD vocoder (d) a male voice using WORLD vocoder.}
  \label{fig:spectrogram}
\end{figure}

 \end{enumerate}
 
\subsection{Subjective listening test}

\begin{figure}
  \centering
  \includegraphics[width=\textwidth]{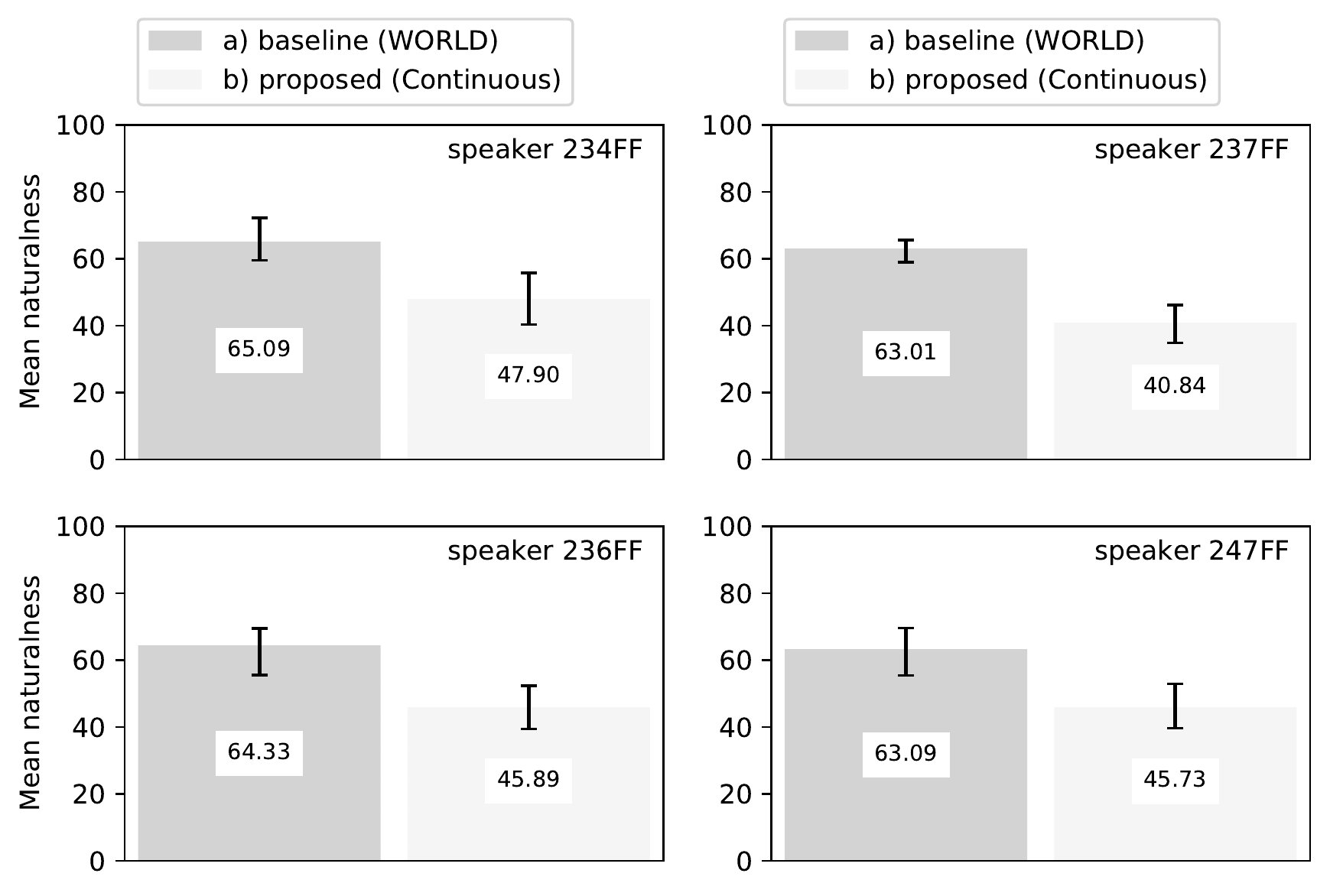}
  \caption{MOS scores for the naturalness question. Higher value means better overall quality. Errorbars show the
bootstrapped 95\% confidence intervals.}
  \label{fig:MOS_results}
\end{figure}

In order to compare TTS versions, we conducted an online MOS-like test. Our aim was to compare the synthesized sentences of the baseline (WORLD) and the proposed (Continuous) vocoders. In the test, the listeners had to rate the naturalness of each stimulus on a scale, from 0 (very unnatural) to 100 (very natural). We chose seven sentences from the CMU-ARCTIC corpus and synthesized them with the four target speakers. The samples appeared in randomized order (different for each listener). Altogether, 56 utterances were included in the MOS test (2 systems x 4 speakers x 7 sentences).
Before the test, listeners were asked to listen to an example to adjust the volume. Each sentence was rated by 11 non-English speakers (2 females, 9 males; 23--39 years old), in a silent environment. On average, the test took 8 minutes to complete. Figure~\ref{fig:MOS_results} shows the average naturalness scores for the tested approaches. For all speakers, the proposed system achieved lower scores than the baseline; and these differences are statistically significant (Mann-Whitney-Wilcoxon ranksum test, with a 95\% confidence level). Although the naturalness scores of the Continuous vocoder did not reach that of the WORLD vocoder, but we can conclude that the speaker adaptation experiment with the proposed vocoder was successful.

\section{Conclusions}
Recent neural vocoders, while capable of high naturalness, also fall short of the criteria for real-time synthesis. In applications requiring low computational complexity, traditional vocoder-based statistical parametric speech synthesis can be beneficial (e.g. a chatbot in a high-load environment like a government website). We use our earlier continuous vocoder in this article, in which the excitation is modeled with two one-dimensional parameters: continuous F0 and Maximum Voiced Frequency. We demonstrate that an average voice can be trained using data from nine speakers, and that speaker adaptation using just 400 utterances is possible (about 14 minutes).

\section{Acknowledged}
The research was partly supported by the European Union's Horizon 2020 research and innovation programme under grant agreement No. 825619 (AI4EU), and by the National Research Development and Innovation Office of Hungary (FK 124584 and PD 127915). The Titan X GPU used was donated by NVIDIA Corporation. We would like to thank the subjects for participating in the listening test.

\bibliographystyle{unsrt}
 \bibliography{mybibliography}

\end{document}